\documentclass[prl,twocolumn,amsmath,amssymb]{revtex4}
\usepackage{graphicx}
\begin{document}
\title{Aging after shear rejuvenation in a soft glassy colloidal suspension:\\ evidence for two different regimes}
\author{F. Ianni$^{1,2}$}
\email{francesca.ianni@phys.uniroma1.it}
\author{R. Di Leonardo$^{2}$}
\author{S. Gentilini$^1$}
\author{G. Ruocco$^{1,2}$}
\affiliation{ $^{1}$ Dipartimento di Fisica, Universit\'a di Roma
``La
Sapienza'', I-00185, Roma, Italy.\\
$^{2}$ SOFT-INFM-CNR c/o Universit\'a di Roma ``La Sapienza'',
I-00185, Roma, Italy. }
\date{\today}

\begin{abstract}
The aging dynamics after shear rejuvenation in a glassy, charged
clay suspension have been investigated through dynamic light
scattering (DLS). Two different aging regimes are observed: one is
attained if the sample is rejuvenated before its gelation and one
after the rejuvenation of the gelled sample. In the first regime,
the application of shear fully rejuvenates the sample, as the
system dynamics soon after shear cessation follow the same aging
evolution characteristic of normal aging. In the second regime,
aging proceeds very fast after shear rejuvenation, and classical
DLS cannot be used. An original protocol to measure an ensemble
averaged intensity correlation function is proposed and its
consistency with classical DLS is verified. The fast aging
dynamics of rejuvenated gelled samples exhibit a power law
dependence of the slow relaxation time on the waiting time.
\end{abstract}
\maketitle
\section{Introduction}
Soft glassy colloidal suspensions are widespread in nature and are
of broad technological importance. Macroscopically, they are
characterized by a non-Newtonian rheology and, often, by a
viscosity that increases many orders of magnitude as time evolves
and the gelation process proceeds \cite{sollich}. These complex
fluids also exhibit a strong sensitivity to external forces and
have mechanical properties typical of soft solids, such as
solid-like behavior below a finite yield stress and thixotropic
response to applied deformation \cite{larson}. In particular, a
shear flow may induce a thinning effect and reduce the fluid
viscosity. At the microscopic level, the gelation process
corresponds to a slowing down of the structural relaxation with
the elapsed time: such a behavior is called aging. Moreover, the
system structural dynamics are accelerated by the shear flow. This
phenomenon is called shear rejuvenation. The competition between
the thickening, produced by the aging, and the thinning, induced
by shear flow, gives rise to
interesting phenomena.\\
On the theoretical and numerical side, evidences of the
competition between these two effects come from the extension of
slow dynamics theories in complex systems, such as mode coupling
theory (MCT) \cite{fuchs}, mean-field models \cite{bbk}, trap
models \cite{sollich} and molecular dynamics simulations
\cite{yamamoto, berthier, angelani}, to the presence of external
driving forces. These studies show that the structural dynamics
are very sensitive to even moderate shear rates, providing that
the characteristic timescale for structural rearrangements becomes
of the same order of the inverse shear rate. In particular, even
starting from non equilibrium, aging states, the presence of shear
ensures the existence of a stationary state whose correlation
function decays to zero on a
timescale given by $\dot{\gamma}^{-1}$.\\
On the experimental side, the investigation of the
shear-influenced slow dynamics at the microscopic level is still
relatively poor. Diffusing Wave Spectroscopy (DWS) \cite{bonnreju,
viasnoff, pascal} and Light Scattering Echo (LSE) experiments
\cite{petekidis, kaloun} gave indirect evidences for a shear
dependent structural relaxation time and for rejuvenation of aged
samples. Unfortunately, the statistical properties of multiple
scattered light (probed in DWS and LSE experiments) are not easily
represented in terms of the particles' correlation functions,
which, on the contrary, is the outcome of the dynamic light
scattering technique in the single scattering regime (DLS). The
DLS technique probes the intermediate scattering function of the
colloidal particles $F_q(t)$ \cite{berne}, which plays a central
role in both theoretical and numerical approaches. The
intermediate scattering function of a colloidal suspension
exhibits two decays: a fast relaxation, which accounts for single
particle diffusion, and a slow relaxation, whose characteristic
timescale $\tau_s$ grows many orders of magnitude with the waiting
time $t_w$ and accounts for the structural rearrangements of the
system. At long $t_w$ the system enters a glass or gel phase; only
the fast relaxation remains and is observed as a decay towards a
plateau, reflecting the
onset of structural arrest.\\
Unfortunately, also DLS suffers of limitations. Indeed, it is not
the proper technique to investigate the system dynamics when the
characteristic slow relaxation time $\tau_s$ becomes very large or
the aging dynamics proceed very fast. The correlation function is
usually measured as an average on the time origin and the
experimental acquisition time needed to get a good signal to noise
ratio must be kept longer than $\tau_s$ and shorter than the time
one should wait before changes in $\tau_s$ are significant. On the
contrary, Multi-Speckle Dynamic Light Scattering (MS-DLS) or
Multi-Speckle Diffusive Wave Spectroscopy (MS-DWS) perform an
ensemble average over the speckle pattern of the light scattered
intensity and acquisition time is strongly reduced. This enables
the investigation of the system dynamics for much longer waiting
times $t_w$ or for faster aging processes. However, in both
techniques, the time resolution is much smaller than in DLS, as it
is limited by the camera device used as detector. Overall, the
experimental studies of aging and rejuvenation under shear of
colloidal suspensions still suffer of technical limitations and
the "ideal" technique has not yet been introduced.\\
On the samples' side, among others, proper candidates for the
study of the aging and rejuvenation phenomena in soft glassy
colloids are suspensions of charged anisotropic colloidal
particles such as clay, which have been widely investigated both
for their important industrial application \cite{olphen} and as a
prototype of soft glassy materials \cite{kroon, bonnaging,
barbara0, barbara1}. Screened charged interactions between these
anisotropic particles induce the formation of disordered arrested
phase in the suspension even at very low volume fractions
\cite{barbara0}. In particular, for sufficiently concentrated
samples, a two stage aging process has been observed through DLS
experiments in a Laponite sample, a suspension of $25$ nm diameter
discoidal charged particles \cite{lequeux}: for small $t_w$, the
structural relaxation time $\tau$ increases exponentially, or more
than exponentially, with $t_w$, while for long $t_w$, when
$\tau>t_w$, a power law behavior shows up. The rejuvenation effect
of a shear flow on such systems has been investigated using DWS
\cite{bonnreju, pascal}, LSE \cite{kaloun}, and DLS \cite{noi}.
The particles' dimension, the sample's history and the kind of
shear applied vary among these works; consequently, the effects of
the rejuvenation on the aging dynamics changes and a comprehensive
picture of the phenomenology has not yet been reached. In Ref.
\cite{noi}, a Laponite suspension is used and the waiting time is
set to zero after the sample filtration. Normal aging in such a
sample is characterized by an intermediate scattering function
having a stretched exponential slow relaxation, whose
characteristic time $\tau_s$ grows exponentially with waiting time
$t_w$ while becoming strongly stretched. Moreover, an exponential
relaxation process is observed at short times and its
characteristic time $\tau_f$ slightly increases with $t_w$
\cite{Lille}. To investigate the evolution of the aging dynamics
during a steady shear, measurements are acquired while the shear
is stopped for a few minutes between different shearing periods.
The system dynamics soon after shear cessation are affected by the
shear flow just applied. In particular, they are characterized by
a strong reduction of aging as soon as $\tau_s$ enters the
timescale set by the inverse shear rate. In \cite{kaloun, pascal},
aged samples of Saponite particles, $125$ nm diameter charged
platelets, are rejuvenated by a strong oscillating shear, after
which the counting of $t_w$ starts. The sample is then submitted
to various periodic shear protocols and dynamics of tracer
particles are investigated during shear and after shear cessation.
Under these conditions, aging is much faster and the structural
relaxation time exhibits a power law dependence on $t_w$.\\
In this paper, we aim to elucidate the apparently contrasting
behavior observed in the two previously described systems
\cite{noi, Lille} and \cite{kaloun, pascal}. We investigate an
aqueous suspension of Laponite, a synthetic layered silicate
composed of monodisperse discoidal particles, which get charged in
water. Aging dynamics following the shear rejuvenation before the
sample gelation occurs are investigated through DLS technique. On
the contrary, aging after rejuvenation of gelled samples is very
fast and classical DLS could not be used, as the system wouldn't
be stationary during the acquisition time. Thus, a novel protocol
of DLS acquisition, consisting in an ensemble average over many
rejuvenating experiments, is proposed to measure the intensity
correlation function and its coherence with classical DLS is
checked. This new method has the double advantage of probing the
intermediate scattering function of the colloidal particles as in
classical DLS, with the "ensemble" average enabling the
investigation of a dynamics changing very rapidly in the system,
while a time resolution much higher than in MS-DLS technique is
achieved. As a result, we find the existence of two different
regimes of aging after shear rejuvenation as a common feature of
charged discoidal clay suspensions. If the shear is applied before
sample gelation is completed, the rejuvenation is followed by a
slow aging regime, behaving just like normal aging after sample
preparation. On the contrary, when a gelled sample is rejuvenated
by shear, we observe a fast aging regime after the shear
cessation, characterized by a power law dependence of the slow
relaxation time on the waiting time.
\section{Experiments and results}
The investigated system consists of an aqueous suspension of
Laponite RD, a synthetic layered silicate provided by Laporte Ltd.
Particles are disk shaped with a diameter of $25$ nm and $1$ nm
thickness. Laponite powder is dispersed in ultrapure water at
$3\%$ wt concentration and stirred for $\sim30$ min. The obtained
suspension, which is optically transparent and initially "liquid",
is loaded into a home made, cone and plate shear cell having a
flat optical window as the static plate. Cell loading (through a
$0.45$ $\mu$m filter) is taken as the origin of waiting times for
normal aging. Incident laser beam (diode pumped solid-state laser,
$\lambda=532$ nm, $P=150$ mW) and scattered light pass through the
same optical window. The scattered light is collected by a
mono-mode optical fiber and detected by a photomultiplier, after
being optionally mixed with a coherent local oscillator field.
Photocounts are acquired through a general purpose, counter/timer
PCI board (National Instruments PCI 6602). We developed a set of
software classes (implemented as extension modules of the object
oriented language Python) designed to perform basic tasks for the
statistical analysis of digital pulse trains. A typical
application is real time multi-tau photon correlation. However, a
software approach, having access to the full photocounts train,
allows to efficiently prototype different analysis protocols,
going far beyond the simple autocorrelation function
\cite{photonlab}. The scattering geometry is fixed (scattering
vector $q=22\;\mu\textrm{m}^{-1}$). The shift of the cell allows
us to select the position of the scattering volume in the cell gap
and the possibility of choosing an heterodyne correlation scheme
\cite{hetero} enables direct access to the detailed velocity
profile in the shear cell.
\subsection{Rejuvenation before gelation}
\begin{figure}
\begin{center}
\includegraphics[height=6.5 cm]{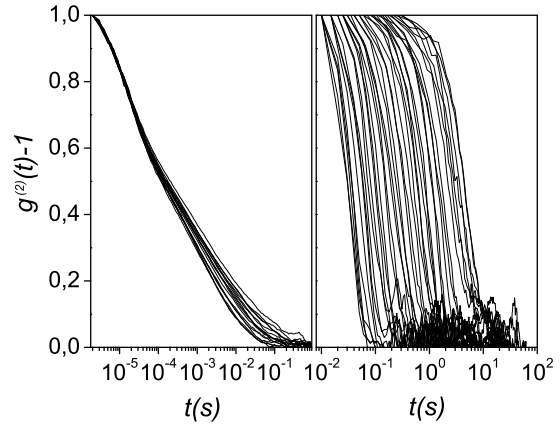}
\caption{Normalized intensity autocorrelation functions for two
different aging regimes after shear rejuvenation. \textit{Left}: a
Laponite sample rejuvenated before entering the arrested phase.
Correlation functions are obtained from an average on the time
origin for 16 equally spaced waiting times between 0.2 and 3 hours
(from left to right) after shear cessation. \textit{Right}: a
gelled Laponite sample for a set of waiting times between $0.3$ s
and $40$ s (from left to right) after shear cessation. Correlation
functions are obtained from an ensemble average over bunches of
counts, all acquired after the application of the same shear rate
for a duration of two minutes.}\label{corrs}
\end{center}
\end{figure}

We first investigate the aging dynamics that follows a shear
rejuvenation induced before the full gelation of the sample takes
place. We let the system age for a time $t_w^0$ after cell
loading, then we apply a shear and finally we follow the dynamics
after shear cessation through DLS. We defining $t_w$ the time
elapsed since shear cessation. Aging of the system is monitored
through the normalized intensity autocorrelation function
$g^{(2)}(t_w, t)=\langle I(q,t_w)I(q,t_w+t) \rangle_T/\langle I(q,
t_w)\rangle_T^2$, where $\langle..\rangle_T$ indicates temporal
average over the acquisition time $T$. In the single scattering
regime and within the Gaussian approximation
$g^{(2)}(t_w,t)=1+|F_q(t_w,t)|^2$ \cite{berne}, where
$F_q(t_w,t)=\langle \rho_{-q}(t_w)\rho_q(t_w+t) \rangle/\langle
\rho_{-q}(t_w)\rho_q(t_w)\rangle$ is the intermediate scattering
function of the colloidal particles. An example of the results of
this procedure is reported in Fig. \ref{corrs}, left panel. In
this example, at $t_w^0=13.4\; h$ a shear rate of $100\;s^{-1}$ is
applied for two minutes; once the shear is stopped, aging is
followed through the intensity autocorrelation function
$g^{(2)}(t_w,t)$ for a set of $t_w$ between 0.2 and 3 hours.
\begin{figure}
\begin{center}
\includegraphics[height=6.5 cm]{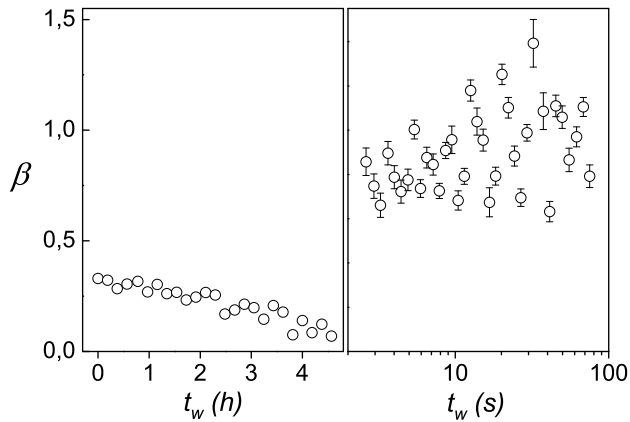}
\caption{Difference of the two aging regimes after shear
rejuvenation as evidenced from the evolution of the $\beta$
parameter, deduced from a stretched exponential fit
$exp[(-t/\tau_s)^{\beta}]$ of the intensity autocorrelation
function. \textit{Left}: aging of a sample rejuvenated before
entering the arrested phase. \textit{Right}: aging of a
rejuvenated gelled sample.}\label{twobeta}
\end{center}
\end{figure}
\begin{figure}
\begin{center}
\includegraphics[height=6.5 cm]{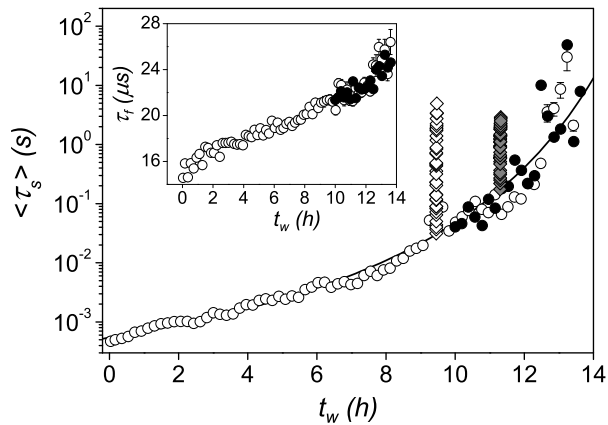}
\caption{Complete rejuvenation of the Laponite sample in the slow
aging regime, \textit{i.e.} before gelation. Average slow
relaxation time $\langle \tau_s\rangle$ is plotted as a function
of waiting time $t_w$ for a sample aging soon after filtration
(empty circles) and for the same sample aging after the
application of a shear rate of $100\:\: s^{-1}$ for 2 minutes,
13.4 hours after filtration (full circles). $t_w$ is shifted in
order to superimpose the first $\langle \tau_s\rangle$ measured
after the application of shear with the first aging curve (empty
circles). The black line is a guide for the eye. In the
\textit{inset} the fast relaxation time $\tau_f$ is plotted as a
function of $t_w$ for the two aging evolution with the same shift
in $t_w$. The slow aging regime is also compared to the fast aging
regime, which is observed after the shear rejuvenation of a gelled
sample: white diamonds represent the evolution of the slow
relaxation time after a shear rate $\dot{\gamma_1}=0.5\:\:
s^{-1}$, while grey diamonds are for $\dot{\gamma_1}=100\:\:
s^{-1}$. Both curves are shifted in $t_w$ through the procedure
previously described.}\label{reju1}
\end{center}
\end{figure}
As for normal aging, the two step decay for $F_q(t_w,t)$:
\begin{equation}
F_q(t_w,t)=f\exp\left[-(t/\tau_s)^\beta\right]+(1-f)\exp\left[-t/\tau_f\right]
\end{equation} where all parameters ($f$, $\tau_s$, $\beta$, $\tau_f$) depend
on $t_w$, provides a very good fit for all the correlations. As
the waiting time $t_w$ evolves, the slow relaxation decay becomes
more stretched: the stretching parameter $\beta$ is smaller than
one and decreases with $t_w$, as shown in the left panel of Fig.
\ref{twobeta}. In order to quantify the slow decay timescale, we
calculate the average slow relaxation time $\langle\tau_s\rangle$
($\langle\tau_s\rangle=\int_0^\infty\exp[-(t/\tau_s)^\beta]\;dt=\tau_s/\beta\;\Gamma(1/\beta)$,
where $\Gamma$ is the Euler Gamma function) and we plot it as a
function of $t_w$ in Fig. \ref{reju1} (full circles). The
evolution of $\langle\tau_s\rangle$ as a function of $t_w^0$ for
the normal aging is also plotted (open circles). In order to show
that shear application completely rejuvenates the system, the
$t_w$ axis for aging after shear rejuvenation has been shifted by
superimposing the first point of the curve on the normal aging
curve. The same behavior can be observed in the evolution of the
fast relaxation timescale $\tau_f$, which is plotted in the inset
of Fig. \ref{reju1}. In fact, the dependence of the fast
relaxation time on the waiting time is in disagreement with
previous observations \cite{bonnaging, lequeux}, that assigned the
fast dynamics to the single particle diffusion.\\
Summarizing, once a shear is applied before gelation, a reduced
timescale for relaxation is observed soon after shear stops. This
starting value depends on the applied shear rate, as evidenced in
Ref. \cite{noi}. The following evolution of the relaxation times
(both fast and slow) traces the same aging curves (once the
waiting time axis is shifted) exhibited during normal aging,
showing that complete rejuvenation is achieved through shear flow.
\subsection{Rejuvenation after gelation}
\begin{figure}
\begin{center}
\includegraphics[height=6.5 cm]{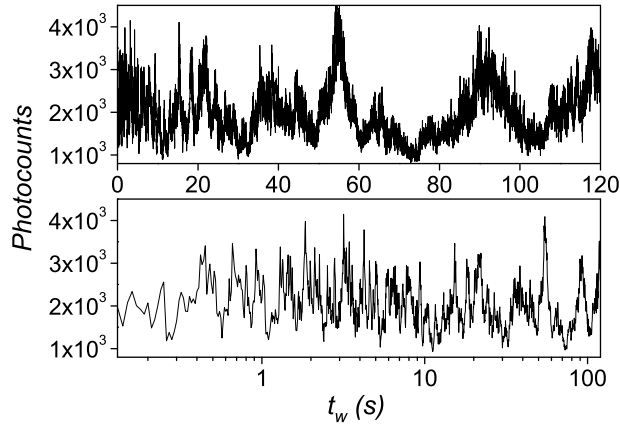}
\caption{Fast slowing down of the scattered intensity fluctuations
with the waiting time for a rejuvenated gelled Laponite sample.
\textit{Top}: the evolution of the counts revealed by the
photomultiplier in $10^{-2}\:\:s$ soon after shear cessation is
plotted, $t_w=0$ corresponds to the shear stop. \textit{Bottom}:
log-linear plot of the counts evolution; a logarithmic binning of
$t_w$ has been performed and the counts have been averaged in each
bin.}\label{counts}
\end{center}
\end{figure}

\begin{figure}
\begin{center}
\includegraphics[height=6.5 cm]{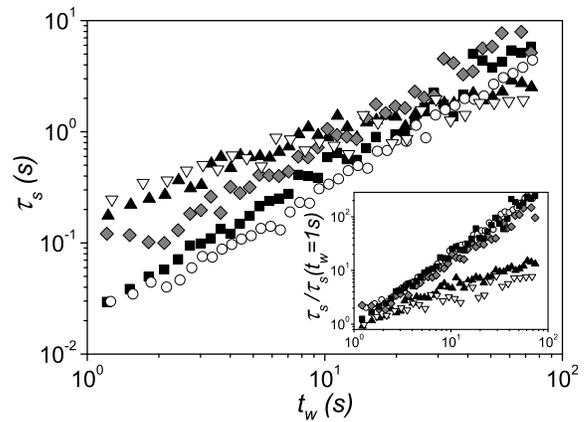}
\caption{Aging after rejuvenation of a gelled Laponite sample. The
evolution with $t_w$ of the slow relaxation time $\tau_s$,
obtained from a single exponential fit of the intensity
auto-correlation function, is plotted for various applied shear
rates $\dot{\gamma}_1$. Circles: $\dot{\gamma}_1=0.5\;s^{-1}$,
squares: $\dot{\gamma}_1=3\;s^{-1}$, diamonds:
$\dot{\gamma}_1=16\;s^{-1}$, down-triangles:
$\dot{\gamma}_1=70\;s^{-1}$ and up-triangles:
$\dot{\gamma}_1=100\;s^{-1}$. All the curves are well fitted by
the power law $\tau_s=A\;t_w^c$. When small enough shear rates are
applied ($\dot{\gamma}_1<<100\;s^{-1}$) the $c$ exponent lies in
the interval $1.23\pm 0.08$. For higher shear rates
($\dot{\gamma}_1=70,100\;s^{-1}$), the $c$ exponent drops to a
value smaller than one, lying in the interval $0.56\pm 0.08$. On
the contrary, the $A$ parameter increases with the shear rate for
all the curves. In the inset, $\tau_s/[\tau_s(t_w=1\;s)]$ is
plotted as a function of $t_w$ in order to better distinguish
among the two groups of curves.}\label{tauvstw}
\end{center}
\end{figure}

\begin{figure}
\begin{center}
\includegraphics[height=6.5 cm]{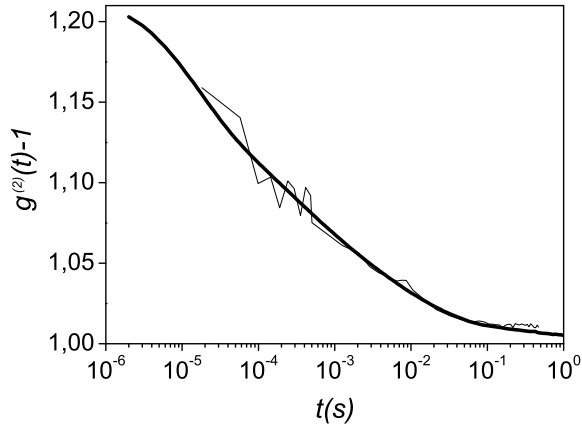}
\caption{Intensity correlation function measured as an ensemble
average over many intensity acquisitions after shear cessation
(thin line) compared to the time averaged correlation function
measured through classical DLS (bold line). Measurements are
performed on a Laponite sample aging under shear at
$\dot{\gamma}=100\;s^{-1}$ for 24 h. As the sample ages during
this long acquisition time, the correlation function plotted in
bold line is an average of all the correlation functions acquired
during the experiment through classical DLS.}\label{confronto}
\end{center}
\end{figure}

We now turn to the investigation of the aging evolution after a
shear is applied to gelled samples. Before describing the
measurements in such a regime, we want to point out that when the
gelled Laponite suspension is put under shear, drastic wall slip
takes place and all the fluid rotates as a solid body leaving a
null shear in the core. Measurements of the velocity profile are
performed through the heterodyne dynamic light scattering setup
\cite{hetero}. In order to apply a controlled shear to rejuvenate
the gelled sample, the solid band is first broken through the
application of an high shear rate ($\dot{\gamma}_0>100\;s^{-1}$)
of the duration of two minutes. We checked that this strong shear
doesn't alter the nature of aging dynamics after subsequent shear
applications. In such a regime, the sample is found to quickly age
back to a gel state. By looking at the intensity scattered by the
sample after shear cessation, the signal fluctuations exhibit a
very rapid slowing down, as can be observed in Fig. \ref{counts}.
Now, being $I(t)$ not stationary for a time period long enough to
collect the data, the intensity correlation function cannot be
obtained by time averaging and an ensemble averaging over many
rejuvenating experiments is used: $g^{(2)}(t_w, t)=\langle
I(q,t_w)I(q,t_w+t) \rangle_e/\langle I(q, t_w)\rangle_e^2$, where
$\langle..\rangle_e$ indicates the ensemble average over several
intensity evolutions acquired after cessation of a repeated shear
application. In particular, after the system has been left aging
for 72 hours in order to reach gelation, a strong shear rate
$\dot{\gamma}_0$ is first applied in order to eliminate wall slip
in the following shear application, then the system is sheared at
a given $\dot{\gamma}_1<\dot{\gamma}_0$ for two minutes. After
shear cessation the intensity fluctuations are collected for a
time interval of two minutes with a time resolution of $.01\:\:s$;
then a shear rate of the same value $\dot{\gamma}_1$ is applied
again for two minutes and the cycle starts again. The whole
measurement lasts several hours and we obtain an ensemble of
hundreds of acquisitions after the same initial conditions are
imposed. In any acquisition, the intensity fluctuations look
stationary in a log-linear plot of the intensity evolution, as
shown in the bottom panel of Fig. \ref{counts}. To speed up the
computation of the correlation function, the acquired counts are
thus logarithmically binned: a logarithmic binning of $t_w$ is
performed and the counts are averaged in each bin. The intensity
autocorrelation function is then calculated as an ensemble average
over all the bunches of counts, in the time window $10^{-2}-10^2$
for a set of waiting times between $1$ and $40$ s (Fig.
\ref{corrs}, right frame). The time interval $t_w<1\;s$ is not
considered, as the correlation functions may be influenced by
inertial effects due to flow stop. In order to show that an aging
of the sample under shear \cite{noi} is negligible during the
whole experiment, we also calculate the correlation functions by
taking only the first or last group of acquired counts and obtain
the same results. The ensemble averaged correlation functions are
first fitted by a stretched exponential decay, the fast component
of dynamics being below the present time window. Results of the
data analysis are checked to be invariant under different binning
and fitting procedures. Compared to normal aging, the new aging
regime entered by the rejuvenated gelled sample is characterized
by correlation functions of a different form and by a much more
rapid evolution of the slow relaxation time with $t_w$. In
particular, the stretching parameter of the correlation function
statistically fluctuates around one and doesn't seem to depend on
the waiting time (Fig. \ref{twobeta}, right panel). The
correlation function can thus be well fitted by a single
exponential decay. Besides, through the effect of the shear, the
structural relaxation time soon after shear cessation is reduced
by many orders of magnitude from the value characterizing the
arrested phase, which must be higher than $100\;s$ (Fig.
\ref{reju1}). Then, $\tau_s$ shows a power law dependence on
$t_w$: $\tau_s=A\;t_w^c$ (Fig. \ref{tauvstw}). To compare the fast
aging evolution in this regime to the slow aging evolution
characterizing the regime before gelation, we added in Fig.
\ref{reju1} the plot of the slow relaxation time evolution in this
second regime, for the applied stresses
$\dot{\gamma}_1=0.5,100\;s^{-1}$. A power law behavior of the
structural relaxation time versus $t_w$, with exponent close to
one, has already been observed after rejuvenation of aged samples
through MS-DWS experiments on clay suspensions \cite{pascal,
kaloun} and on other colloidal glasses \cite{viasnoff}. Also in
normal aging, clay suspensions have shown such a power law
behavior, but only for very long waiting times
\cite{lequeux}.\\
In order to investigate the power law behavior of the aging
dynamics as a function of the shear rate value applied for
rejuvenation, we repeated the same experiment by varying
$\dot{\gamma}_1$. For small applied shear rates, Laponite aged
samples exhibit a nonlinear velocity profile \cite{bonnrheo}.
Heterodyne photocorrelation is thus used to measure the detailed
velocity profile, in order to calculated $\dot{\gamma}_1$ as the
local shear rate in the scattering volume \cite{berne}. On the
contrary, for higher applied shear, the shear rate is constant
along the gap and $\dot{\gamma}_1$ can simply be calculated as the
cone velocity over the gap width. For
$\dot{\gamma}_1<<\dot{\gamma}_0$, no significant changes are found
in the $c$ exponent of the power law. Indeed, for different values
of $\dot{\gamma}_1$, spanning more than one decade:
$\dot{\gamma}_1=0.5,3,16\;s^{-1}$, all the aging curves are well
fitted by the power law with $c$ lying in the interval $1.23\pm
0.08$ (Fig. \ref{tauvstw}). When $\dot{\gamma}_1=70,100\;s^{-1}$
instead, the $c$ parameter drops to a value smaller than one and
lies in the interval $0.56\pm 0.08$. On the contrary, for all the
shear rates here investigated, the $A$ parameter increases with
$\dot{\gamma}_1$, but it is not clear why the relaxation time soon
after shear cessation is longer for an higher applied shear rate.
In particular, one would expect that when the shear rejuvenation
effect dominates aging, the inverse shear rate sets the slow
relaxation timescale. However, in our data, we don't find this
scaling for the slow relaxation time soon after shear cessation.
For example, at small shear rates, the shear does accelerate the
system dynamics, but the slow relaxation time gets much smaller
than the inverse shear rate.\\
Finally, consistency with classical DLS has been checked for this
original method of DLS employed to measure the intensity
correlation function for a rapidly aging sample. In order to
compare the correlation functions measured through both methods,
we monitored the aging dynamics of a Laponite sample under shear
at small $t_w^0$. In this regime, the system dynamics after shear
cessation are stationary over the timescale of the tens of seconds
and can thus be investigated through classical DLS also. Soon
after cell loading, the sample is left aging under shear at
$\dot{\gamma}=100\;s^{-1}$ and shear is stopped each $60$ s, a
pause of $5$ s is taken in order to avoid inertial effects due to
flow stop and then the scattered intensity is acquired for $1$ s,
with a time resolution of $10^{-5}$ s. Each hour, the
time-averaged intensity correlation function is also measured
through classical DLS by stopping the shear for $80$ s. The
experiment lasts 24 h and the ensemble averaged intensity
correlation function is calculated from all the acquired bunches
of counts. Such a long acquisition time is necessary in order to
reach a good statistics in the correlation function. In comparison
to MS-DLS technique, where the intensity correlation function is
calculated as an ensemble average over the speckle pattern, the
acquisition time is extremely large, but the time resolution is
much higher and enables the investigation of faster dynamics. As
expected for this regime, the intensity correlation function is
independent on $t_w$, the time elapsed since shear stop. On the
contrary, as the system under shear keeps on aging with $t_w^0$
\cite{noi}, the dynamics change during the experiment and the
ensemble averaged correlation function will provide an average
value of the slow relaxation timescale. The ensemble averaged
intensity correlation function is thus compared to the average of
all the intensity correlation functions measured during the
experiment through classical DLS (Fig. \ref{confronto}). In
conclusion, intensity correlation function measured through an
ensemble average over many acquisitions of the intensity evolution
after shear cessation shows a good agreement with the one measured
through classical DLS, where an average over the time origin is
performed.
\section{Conclusions}
We show the existence of two different regimes of aging for a
suspension of charged discoidal clay particles after shear
rejuvenation. Before gelation, classical DLS measurements have
shown that a shear flow can completely rejuvenate the sample. The
system dynamics after shear cessation depend on the applied shear
rate and the following aging evolution is identical to the one
after sample preparation: the fast relaxation time increases
slightly with the waiting time $t_w$, while the slow relaxation
time grows exponentially with $t_w$ and becomes strongly
stretched. On the contrary, in a gelled sample, aging dynamics
after shear rejuvenation proceed very rapidly, so classical DLS
measurements cannot be used and a novel dynamic light scattering
method to measure the intensity correlation function has been
proposed. An ensemble average over many rejuvenating experiments
is used and the validity of this method is checked through
comparison with classical DLS measurements. With this method, we
investigated the aging dynamics of a gelled sample after shear
rejuvenation by measuring the ensemble averaged intensity
correlation function. Its time evolution is different from the one
typical of the other regime, as its decay is characterized by a
single exponential. Moreover, the slow relaxation time soon after
shear stop is reduced by many orders of magnitude with respect to
the large relaxation time typical of the arrested phase, while a
very rapid evolution of the aging follows. In particular, we
observed a power law behavior of the slow relaxation time with
$t_w$, whose
exponent shows two different values for small or high applied shear rates.\\
Further studies are needed to clarify the nature of these
different aging evolutions after shear rejuvenation and how they
are linked to the gelation process.\\
The authors wish to thank Gianni Bolle and Md Islam Deen for
technical assistance.

\end{document}